\documentclass[12pt]{iopart}

\usepackage{lineno}

\usepackage{amssymb}

\usepackage{txfonts}

\usepackage[pdftex,colorlinks=true,bookmarks=false,citecolor=blue,urlcolor=blue]{hyperref} 

\usepackage{graphicx}
\usepackage{dcolumn}
\usepackage{bm}

\usepackage{multicol} 

\usepackage{CJK}

\begin{document}


\title
{Infinite Maxwell fisheye as a complement for an imaging device}

\author{Yangjie Liu()$^{1,2}$ and Huanyang Chen()$^{2,*}$} 

\address{$^1$ Engineering Building, School of Electronic Engineering and Computer Science, Queen Mary, University of London, 327 Mile End Road, E1 4NS Mile End, London UK}

\address{$^2$ Physics Building, School of Physical Science and Technology, Soochow University, 1 Shizi Street, 215006 Suzhou, P. R. China}

\ead{ \mailto{Correspondence author--$^{*}$chy@suda.edu.cn}}

\begin{abstract}
\noindent This manuscript proposes a new imaging medium via isotropic refractive index of heterogeneous medium. We exploit conformal-map to transfers the full Maxwell fisheye into a mapped profile within a unit circle. The imaging resolution of this new profile can be maintained as good as the mirrored Maxwell fisheye profile or its conformally-mapped counterpart in previous literature.   



\end{abstract}

\footnote {started 4 Jan, finished 20 Jan, revised 6 May, re-revised 29 May 2015, submitting to Special issue on Transformation Optics in {\itshape J. Opt.}}

\maketitle


\section{Introduction}

Continual interest in quest for perfect imaging in optical design, recently reinvigorates the efforts to advance the imaging resolution to cater for human's visual observation or other entertainment requirements. Despite the material-loss-suffering concept of perfect lensing via negative refractive index~\cite{Pendry2000,Pendry2004,Pendry2008,Fu2008,Belov2009}, an alternative approach to achieve subwavelength imaging using positive refractive index was proposed~\cite{Leonhardt2009a}, namely mirrored Maxwell fisheye(a finite spatial-varying profile), which was based on manipulating curved geometry of light~\cite{Leonhardt2009b}. This idea to design the geometry for light {\itshape de facto} gave birth to a novel sub-field of transformation optics~\cite{Leonhardt2006, Pendry2006, Leonhardt2009, Leonhardt2010book,Xu2014b}. A worth-noticing paper~\cite{Tyc2014a} declares that an active drain in a mirrored Maxwell fisheye does not give subwavelength resolution of imaging.  Consequently, an extension to combine coordinate transformation(herein precisely conformal mapping) and spatial profile(herein mirrored Maxwell fisheye profile) showed an interesting duality of simultaneous cloaking and imaging~\cite{Qiannan2013}, but no thorough investigation on the imaging resolution was performed. This manuscript's novelty is two fold: one in treating this resolution aspect in a transformed full Maxwell fisheye, the other in shrinking an infinite Maxwell fisheye profile into a finite unit circle by conformal map~\cite{Leonhardt2006,Xu2014b}. The previous investigation~\cite{Qiannan2013} made use of a mirrored Maxwell fisheye, which necessitated a forbidden cavity region delineated by a curved shell of perfect electric conductor(PEC). An equivalent profile therefore, can also be investigated: the {\itshape full} Maxwell fisheye without this unnecessary shell of PEC. This equivalence relies on the fact that their optical path between the full and the mirrored Maxwell fisheye are equal~\cite{Leonhardt2009a}. Here, we propose another heterogeneous medium for light to perform similar functionality of good imaging, as proposed in~\cite{Qiannan2013}.

\section{Profile design for proposed imaging}\label{profile}
\begin{figure*}[ch]
\includegraphics[width=0.7\textwidth]{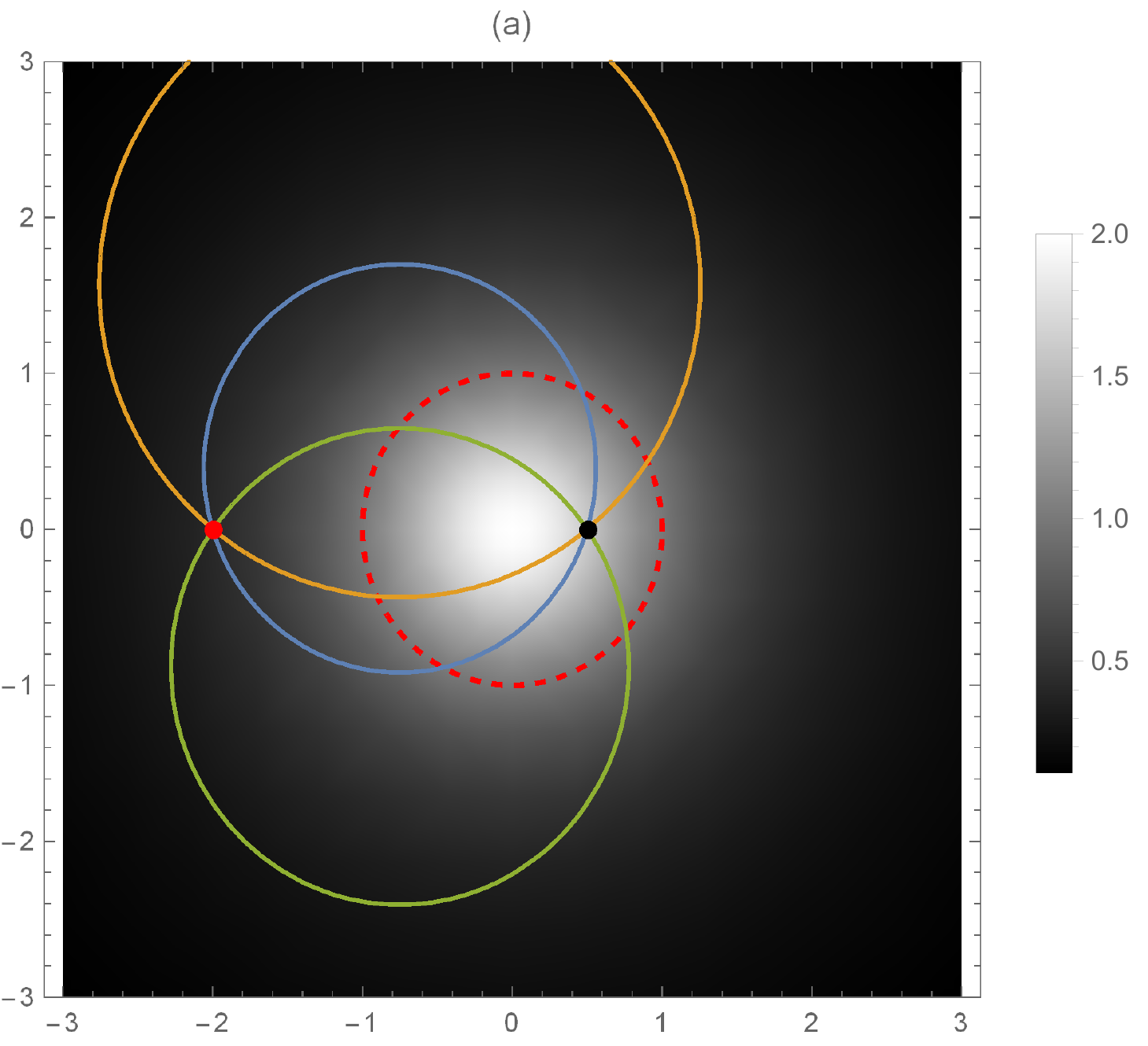}\\
\includegraphics[width=0.7\textwidth]{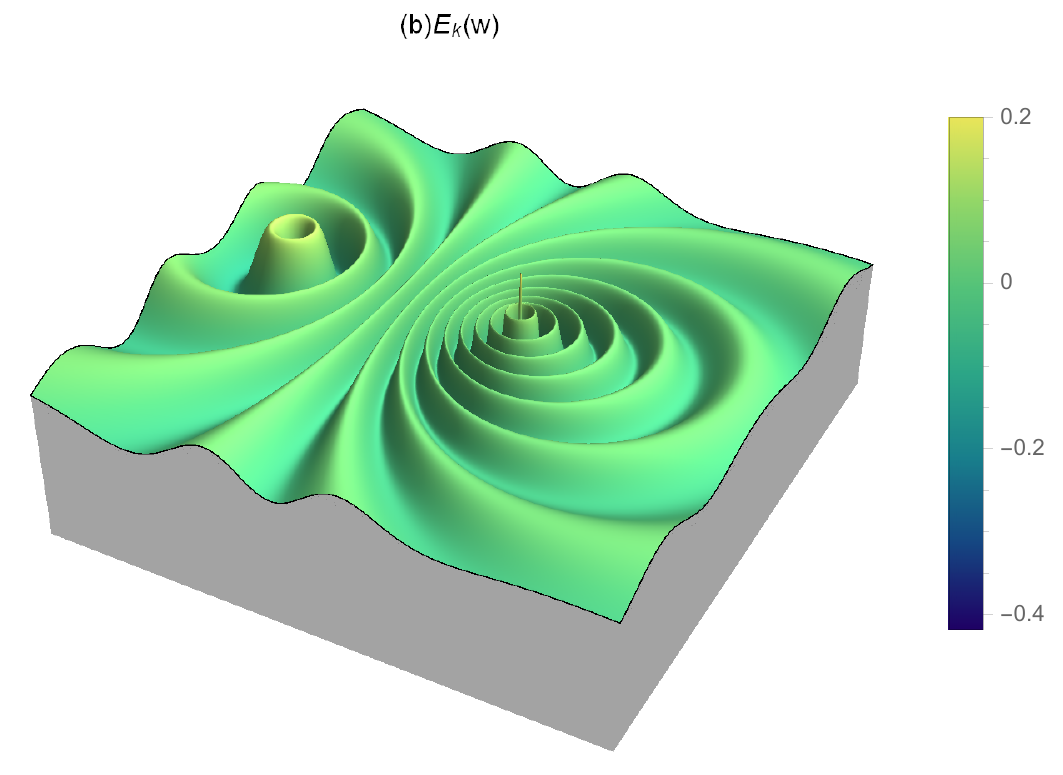}
\caption { \label{fig:Fig1} (Color online) (a) Ray-tracing diagram for a full Maxwell fisheye(focus (0.5,0) originated from source (-2,0)), in which red dashed circle delineate at $r=r_w$ in a full Maxwell fisheye; (b) Wave value distribution $E_k(w)$ in arbitrary unit corresponding to (a) according to \cite{Leonhardt2009b}. Parameters for (a-b): image point$(0.5,0)$, scaling factor $a=1$,fisheye radius$r_w=1$;$l=24,\gamma=\arctan 2, \chi=\pi$ in ~\eref{nxy}, \eref{lambda} in the forthcoming ~\sref{subsec2p1}.}
\end{figure*}

Let us start from the full Maxwell fisheye profile for refractive index, 
\begin{equation}\label{MFE}
n(r)=\frac{2}{1+\big(\frac{r}{r_w}\big)^2}, r\in[0,\infty)
\end{equation} 
where all light rays make circles($r_w$ is a characteristic radius). It was Luneberg who realised that Maxwell fisheye profile can be generated by stereographic projection from a sphere surface to a plane~\cite{Leonhardt2009,Leonhardt2010book}. By the idea that light travels along geodesics along sphere(in virtual space), an optical implementation of imaging device follows: all rays emitted from any point in full Maxwell fisheye will converge at its conjugate point--antipole--to form an image point~\cite{Leonhardt2009a}(cf. converging circles in figure~\ref{fig:Fig1}(a)). An obvious fault for the full profile of 2D Maxwell fisheye is its infiniteness in area and hence the truncation trick along with mirror was employed~\cite{Leonhardt2009a}. However, here we attempt to sidestep this obstacle by applying conformal-map method~\cite{Leonhardt2006}.

We use the conformal-map to transform imaging behaviour of the full Maxwell fisheye. In principle, one can implement any conformal map to this Maxwell fisheye profile to create a new heterogeneous medium to perform similar imaging functionality as in \fref{fig:Fig1}. A conformal map creates a virtual space where light travels but this virtual space is made of a stack of Riemann sheets in general~\cite{Leonhardt2006,Leonhardt2006b}. In this paper, Zhukovski map is chosen which creates two Riemann sheets which are connected by their branch cut~\cite{Leonhardt2006,Qiannan2013}. Because the full and mirrored Maxwell fisheye preserve the same optical path(cf. virtual sphere in Figure 3(A)~\cite{Leonhardt2009a}), the mirror in previous work~\cite{Qiannan2013} can be removed so as to dismantle the 
constraint of the curved shell of PEC, by simply taking the full fisheye on the lower sheet of virtual space. Equivalent to ray-tracing in \fref{fig:Fig1}(a), one obtains a wave field picture $E_k(w)$~\cite{Leonhardt2009a} on the lower Riemann sheet (half of the virtual uniform space) in figure \ref{fig:Fig1}(b). 

\subsection{Conformally-mapped Maxwell fisheye}\label{subsec2p1}

\begin{figure}[h]
\includegraphics[width=1.2\textwidth]{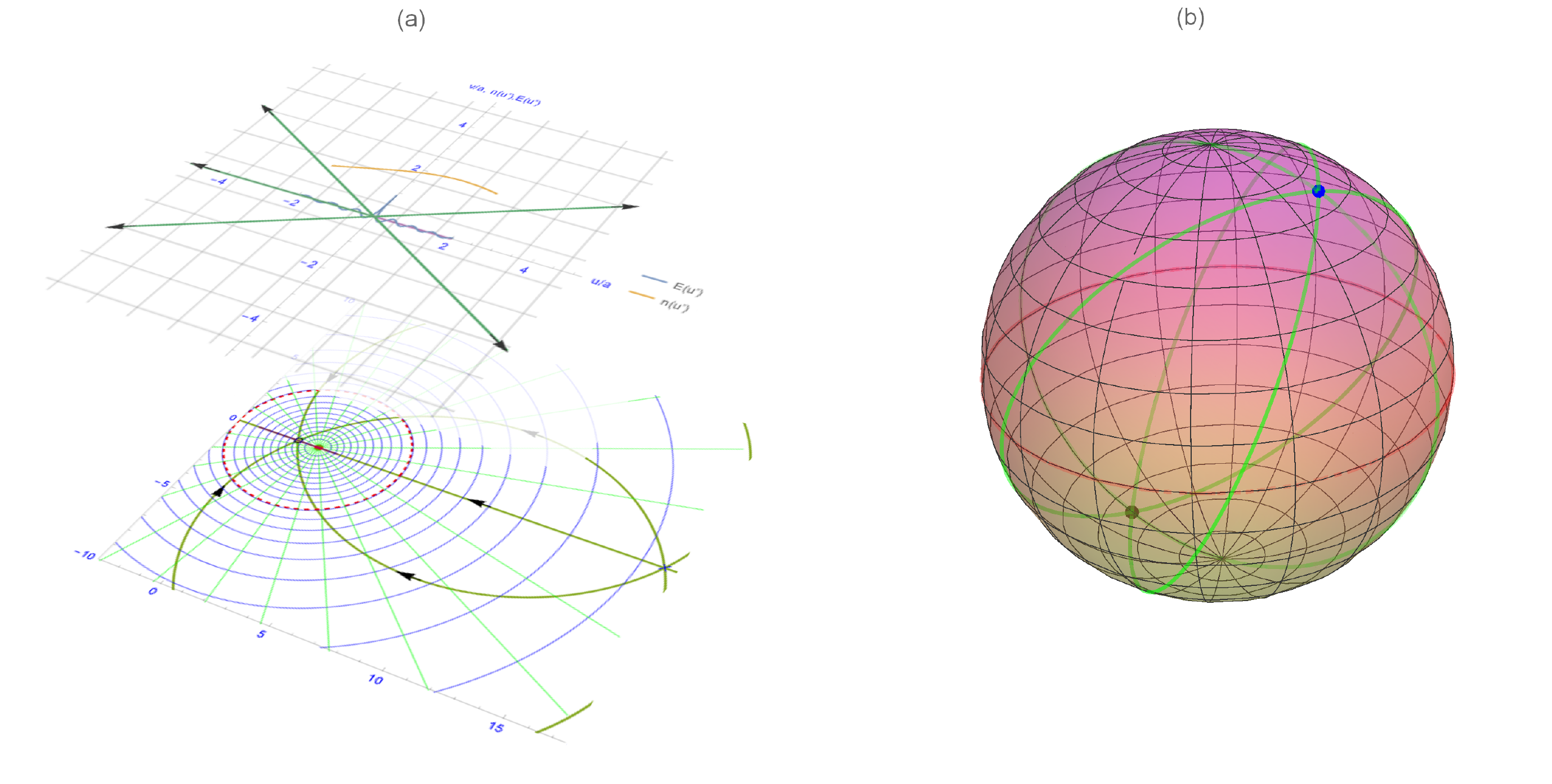}

\caption {\label{fig:Fig3} (Colour online) (a) Illustration of the lower $(u',v')$ and the upper $(u,v)$ Riemann sheet. On the lower sheet, red dashed circle indicates the coordinate $r_w=1$, blue contours for $\theta=\arccos z/r_w\in[0,\pi)$ and green for $\phi\in[0,2\pi)$ as curved coordinates. The red dot $M_w$ indicates the centre of the Maxwell fisheye put on the lower sheet. Thick green circles(including a straight line through the branch cut in purple) are three light trajectories contributing to form an image at the origin point in hollow black circle $O_w$ from a source point in blue cross $S_w$: $O_wM_w=b,M_wS_w={r_w^2}/{b}$. Notice that the light trajectories deflect at the branch cut and transmit onto the upper sheet(reflection overlooked). The wave distribution $E(u')$ \cite{Leonhardt2009a} and the refractive index $n(u')$ for the branch cut on the lower sheet are plotted on the upper sheet. Parameters: $b=1,r_w=4,l=24$.
(b) Virtual curved space $(\theta,\phi)$ on a sphere, for the lower sheet in (a) under stereographic projection. On the sphere light travels in geodesics, in green circle, diverging from blue source and converging at black image point(geodesics equation based on Eq.~(19.9)~\cite{Leonhardt2010book} with the corresponding initial conditions). }  
\end{figure}

For a fair comparison with our previously-published wave picture~\cite{Qiannan2013}, we stick to Zhukovski map
\begin{equation}
w(z)=z+\frac{a^2}{z},(a>0)
\end{equation}
in optical conformal mapping method~\cite{Leonhardt2006} although principally any conformal map reserves the imaging behaviour of Maxwell fisheye~\eref{MFE}. Under Zhukovski map, a physical space(plane) is mapped onto two Riemann sheets, of which the lower and upper sheets are geometrically connected by the branch cut--a segment from $-2a$ to $2a$. The two sheets are labelled by two Cartesian coordinates $(u,v)$ and $(u',v')$ shown in \fref{fig:Fig3}(a). 

We choose to shift the centre of Maxwell fisheye profile, $M_w$, away from the origin by an arbitrary distance $b$ on lower sheet. The merit of using conformal map in this case is to transfer imaging point from underneath to above the ground. Therefore one has to put the image point(marked in hollow circle on the lower sheet in figure~\ref{fig:Fig3}(a)) on the branch cut~$[-2a,2a]$ (illustrated by purple segment in figure~\ref{fig:Fig3}(a-b)) and thus the source(marked in blue cross in figure~\ref{fig:Fig3}(a)) away from that. This is because a pair of image and source in a full fisheye appear on opposite sides of the centre(marked in red dot) along a line joining the source and the image(cf. figures~\ref{fig:Fig1} and \ref{fig:Fig3}(a)). If the source point fell on the branch cut, its equivalent in $z$ physical space would degenerate into two sources(which images to two points separately) off $u'$-axis on the unit circle(branch cut), when parameter $b$ is swept over under the condition $r_w\leq a$. In this case, the image point would instead lie underground(in the unit circle) in the virtual space and one feature of conformal imaging~\cite{Chenhy2013}--the double degeneracy of one source point imaging to two points-- is broken. Furthermore, the ray-tracing computation in forthcoming \sref{ray} is doomed as well. This falls beyond the scope of this manuscript and we restrict the fisheye size under the condition $r_w>a$.

Thus based on the optical conformal mapping method~\cite{Leonhardt2006,Leonhardt2010book}, the refractive index profile(in physical space, its range $\sim[0,33]$\footnote{It is interesting to notice that the zero points of refractive index occur at $z=\pm 1$, which aligns with the doom of trajectory-tracing from Hamilton's equations at these two points.}) for our conformally-mapped full Maxwell fisheye is written as
\begin{eqnarray}\label{nxy}
n(x,y)=n_w\Big\vert \frac{{\rm d}w(z)}{{\rm d}z}\Big\vert=\Big\vert \frac{{\rm d}w(z)}{{\rm d}z}\Big\vert\cdot\cases{1,& for $\vert z\vert>a$;\cr
\frac{2}{\big\vert\frac{w-b}{r_w}\big\vert^2+1},& for $\vert z\vert\leqslant a$.}\Bigg\vert{z\equiv x+iy}
\end{eqnarray}
Considering TE polarization where electric field directs perpendicular to the complex $z$ plane, the scalar Helmoltz equation stands for any wavelength~\cite{Leonhardt2009a}:
\begin{equation}\label{lambda}
\lambda =\frac{2\pi r_w}{\sqrt{l(l+1)}}, l>0.
\end{equation}
To compare fairly with previous simulation~\cite{Qiannan2013}, we follow to take integer values of $l$ in this manuscript but $l$ can be non-integer as well, which would vary the imaging phase.  

Notice that the full Maxwell fisheye profile bends the uniform coordinates ($u',v'$ Cartesian)on lower $w$ sheet into curvilinear ones ($\theta,\phi$ spherical) as plotted in lightblue and lightgreen in figure~\ref{fig:Fig3}(a), according to inverse stereographic projection(One can calculate arc tangent value via the command of \texttt{ArcTan[u'-b,v']} in \texttt{Mathematica 10.0}~\cite{Mathematica}, taking into account with which quadrant the point $(u'-b,v')$ is in), 
\begin{eqnarray}
\theta=\arccos\frac{\vert u'-b+iv'\vert^2-r_w^2}{\vert u'-b+iv'\vert^2+r_w^2}&\in[0,\pi],\\
\phi=\arctan(u'-b,v')&\in[0,2\pi).
\end{eqnarray} 
An imaging process consists of the light flow from any source point(marked in blue cross) towards its antipodal(in hollow circle) and then leakage(refraction) onto the upper sheet, illustrated by three representative light paths in thick green curves in figure~\ref{fig:Fig3}(a). It is interesting to notice that all rays emanating from the source do not refract on the upper sheet since total reflection could occur for partial of them. Thus the light trajectories on the lower sheet only occupy a finite area, out of which is the forbidden zone for light.  

\section{Wave simulation}\label{sim}
\begin{figure}[h]
\includegraphics[width=0.5\textwidth]{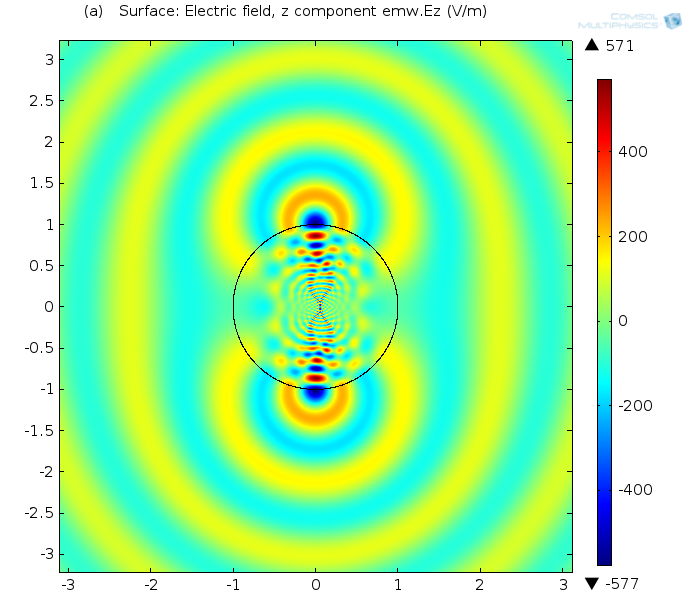}
\includegraphics[width=0.5\textwidth]{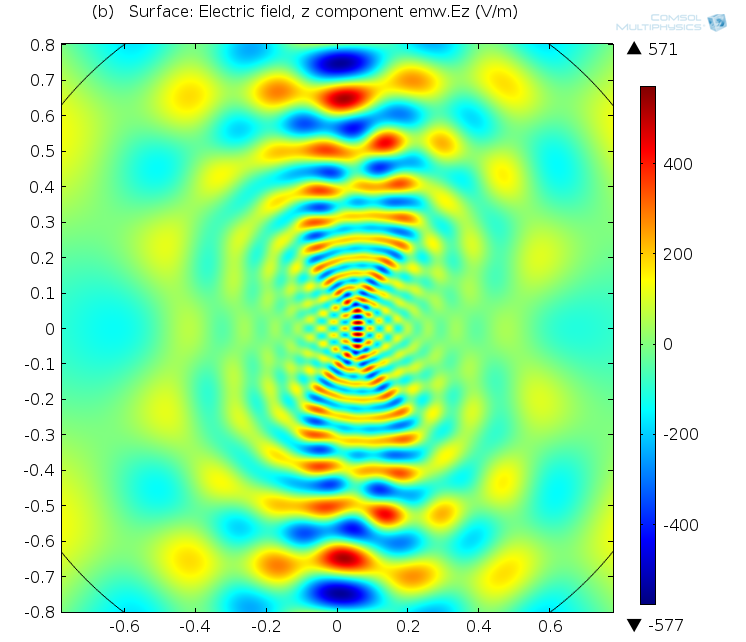}

\caption {\label{fig:Ez} (Colour online) (a)Electric field distribution (in V/m) in $z$ direction(\texttt{COMSOL} simulation); (b)Zoom-in of (a). (a-b) Unit circle is marked black to indicate the boundary between two Riemann sheets in virtual space. Parameters: $b=1,r_w=4,l=24$. }  
\end{figure}

In two-dimensional case, put an active point source($1\rm A$) at $S(z)$ on $z$ plane, we simulate on the TE field distribution($E_z(x,y)$) via \texttt{COMSOL Multiphysics}~\cite{COMSOL}. As a preliminary demonstration, we plot a distribution of TE field in figure~\ref{fig:Ez} under a certain set of parameters. Notice a truncation is taken beyond certain radius away from unit circle(for instance, $10r_w$) since the heterogeneous medium there can be approximated as unity similar to \cite{Qiannan2013}. We choose to position the source so that the corresponding image falls at the middle of the branch cut--the origin in virtual space($\pm i$ in physical space), as \sref{profile} explains. 

This feature of unity refractive index far away from the unit circle, allows one to position two image detector in the free space environment, which may be easy to realise in experiment. The discontinuous pattern at the branch circle $\vert z\vert =1$ can be attributed to the slight jump between their refractive indices of two Riemann sheets. This discontinuity is also present in previous simulation work~\cite{Qiannan2013}(Fig.~2 therein), which may be improvable by designing a layer of gradient index smoothly varying but we stick to the simple profile \eref{nxy} for simplicity. In figure~\ref{fig:Ez}(b), the light originates from source point $S(z)$ inside the unit circle and diverges out on two distinctive paths upward until to the image point $i$ on the upper half plane and $-i$ on the lower half plane as well(mirror-symmetry), similar to the case of mapped mirrored Maxwell fisheye~\cite{Qiannan2013}. 

In order to seek the best resolution of imaging, we sweep over several parameters to investigate on the image resolution with respect to the operating wavelength in accordence with ~\eref{lambda}.

\subsection{Sweeping the centre of the fisheye via $b$}

\begin{figure}[hc]
\includegraphics[width=0.7\textwidth]{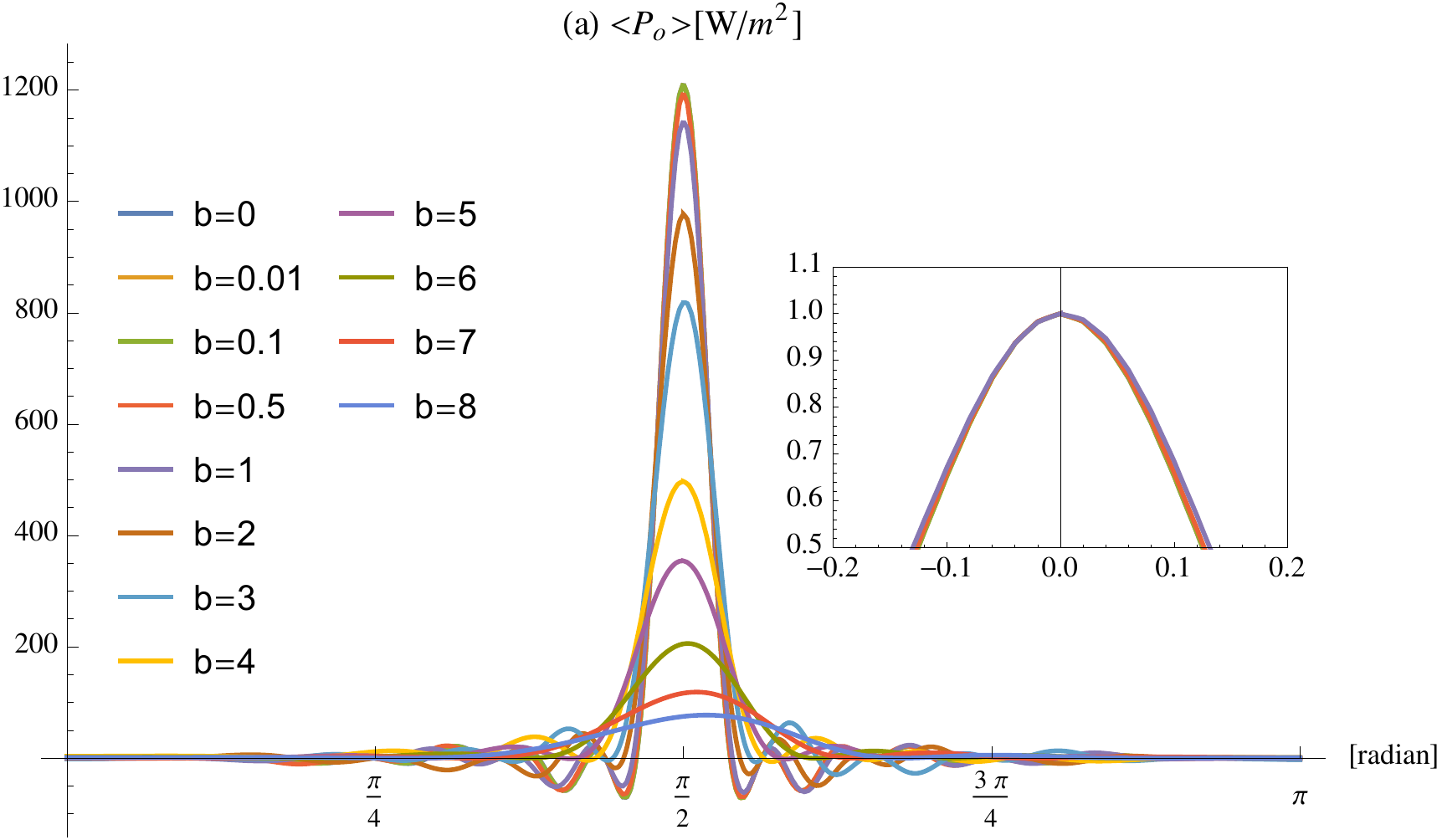}\\
\includegraphics[width=0.5\textwidth]{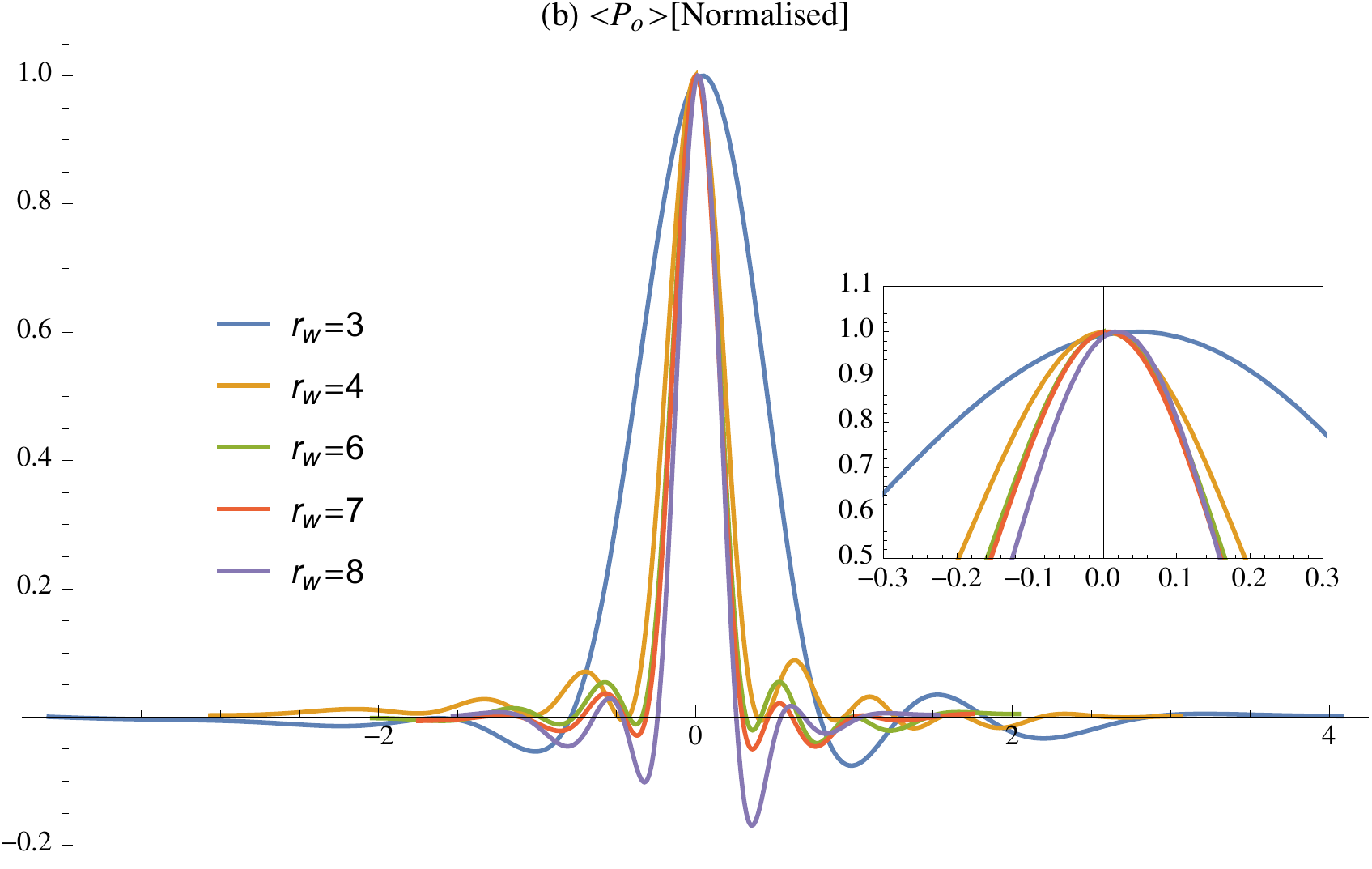}
\includegraphics[width=0.5\textwidth]{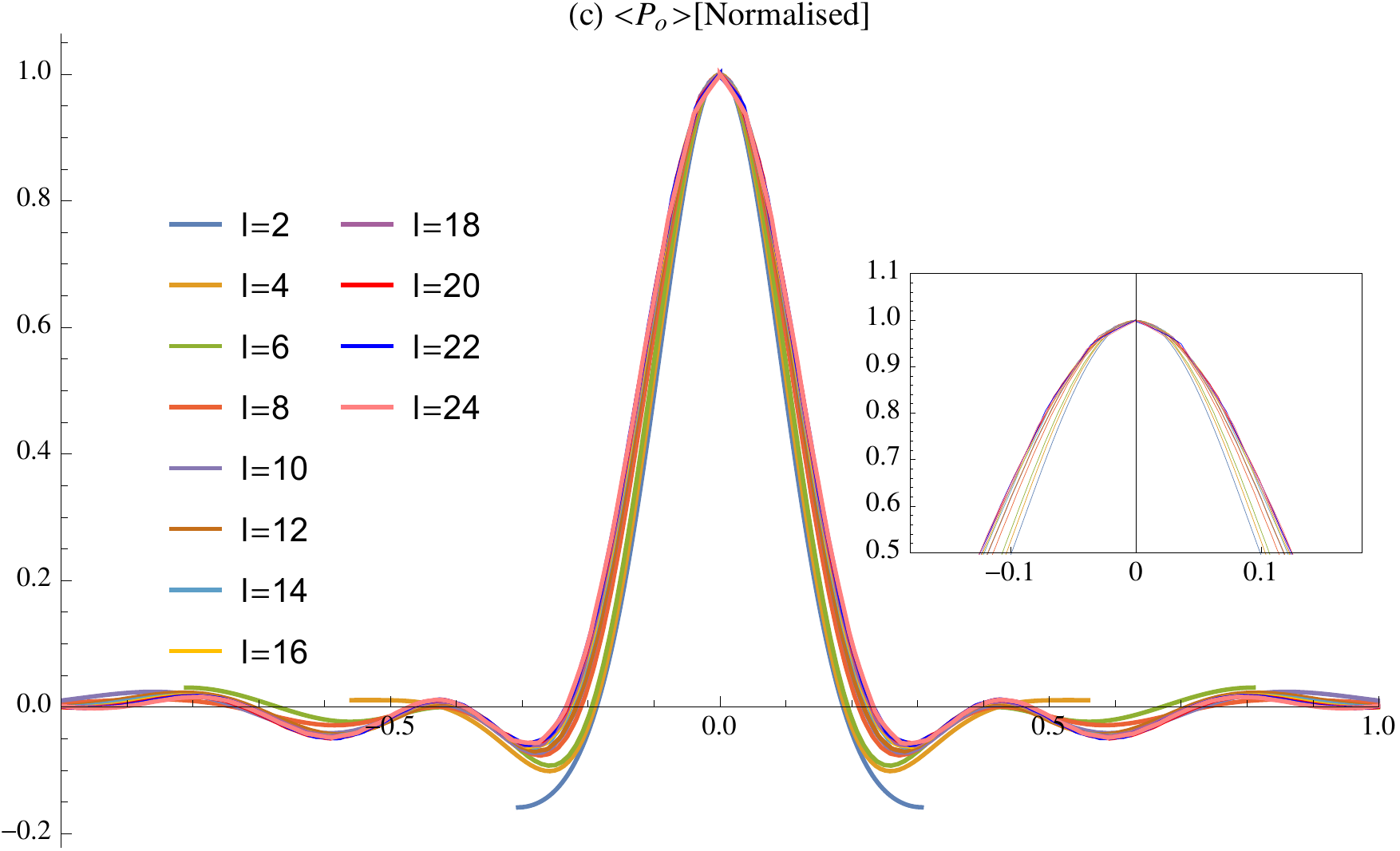}

\caption {\label{fig:sweeping} (Color online) Average energy flowing out along the upper half circle when one sweeps over (a) $b$, (b) $r_w$ and (c) $l$. Note that (b-c) main panels and all three insets are all normalised in ordinate and abscissa to directly compare with the operating wavelength in physical space:$\lambda/n(z=i)=\lambda/2$ according to \eref{nxy} and \eref{lambda}. Parameters: (a)$r_w=4,l=24$(the same wavelength $\lambda=0.5130\rm {m}$ as in ~\cite{Qiannan2013}); (b) $b=4,l=24$; (c) $b=0, r_w=4$. }  
\end{figure}

Firstly, to find an optimal image resolution for the considered profile, we opt to shift the centre of the fisheye along $u'$-axis on the lower $w$ sheet(sweeping $b$). Due to the geometrical symmetry of the case, varying $b$ within positive regime is sufficient to disclose the resolution scaling. In figure~\ref{fig:sweeping}(a), average energy flowing out along the upper half circle(branch cut) is plotted versus radiant angle, under different values of $b$. It is found that as $b$ decreases to zero, the image resolution becomes sharper until a lower limit of around half wavelength if we define the imaging resolution by full width at half maximum(FWHM) as shown in figure~\ref{fig:sweeping}(a) inset.

\subsection {Sweeping the size of the fisheye $r_w$}

Secondly, the geometrical size of Maxwell fisheye on the lower sheet can be varied as well. Although the imaging resolution appears shrinking as $r_w$ increases, it never shrinks below 0.3 wavelength in figure~\ref{fig:sweeping}(b)(in \emph{physical space}, the actual wavelength is $\lambda/n(z=i)=\lambda/2$). The sharper trend with increasing $r_w$ may imply the refractive index jump along unit circle $\vert z \vert =a$, ignored by the previous study~\cite{Qiannan2013}. As $r_w$ increases, the discontinuity in the refractive index $n(z)$ at the unit circle diminishes and thus better resolution is achieved. This discontinuity in the refractive index may also explain the scattered colour pattern in figure~\ref{fig:Ez}.

\subsection{Varying the wavelength via eigenvalue $l$}
A similar trend can be observed when one tunes integer values of $l$ in figure~\ref{fig:sweeping}(c), that the imaging resolution evolves worsen as $l$ increases but still no shaper resolution than 0.2~ wavelength(corresponding to the least value $l=2$) can be achieved herein, which is comparable to the resolution obtained in~\cite{Qiannan2013}. As mentioned in \ref{subsec2p1} about \eref{lambda}, only integer valuesof $l$ are taken but non-integer value applies as well.

To summarise \sref{sim}, we sweep over three parameters to observe the imaging resolution dependence on them. The optimal parameter according to our simulation is: $b=0, r_w=8$ and $ l=2$ and the optimal resolution achievable in our scenario is $0.2\sim$wavelength according to our definition of FWHM.


\section{\label{ray}Ray-tracing diagrams}
The bifurcating behaviour of light wave, inside the unit circle in figure~\ref{fig:Ez} tempts one to wonder the propagation scenario of light there. One simple method is to investigate into the ray-tracing diagrams. When it comes to ray-tracing in virtual space, it is straightforward to predict that the light rays turn out to be circles underneath and straight lines above(cf. \fref{fig:Fig3}(a), if one neglects reflection at the branch cut\footnote{Notice that the discussion for null reflection from an incidence wave from the upper Riemann sheet (on equations (11)-(13), p4~\cite{Chenhy2011}) does not apply since in our scenario the incidence wave comes from the lower sheet instead of from the upper in~\cite{Chenhy2011}. }). However, it may be of interest to peep into ray tracing in physical space in transformed Maxwell fisheye. This section will address this concern. 

\begin{figure}[h]
\includegraphics[width=0.7\textwidth]{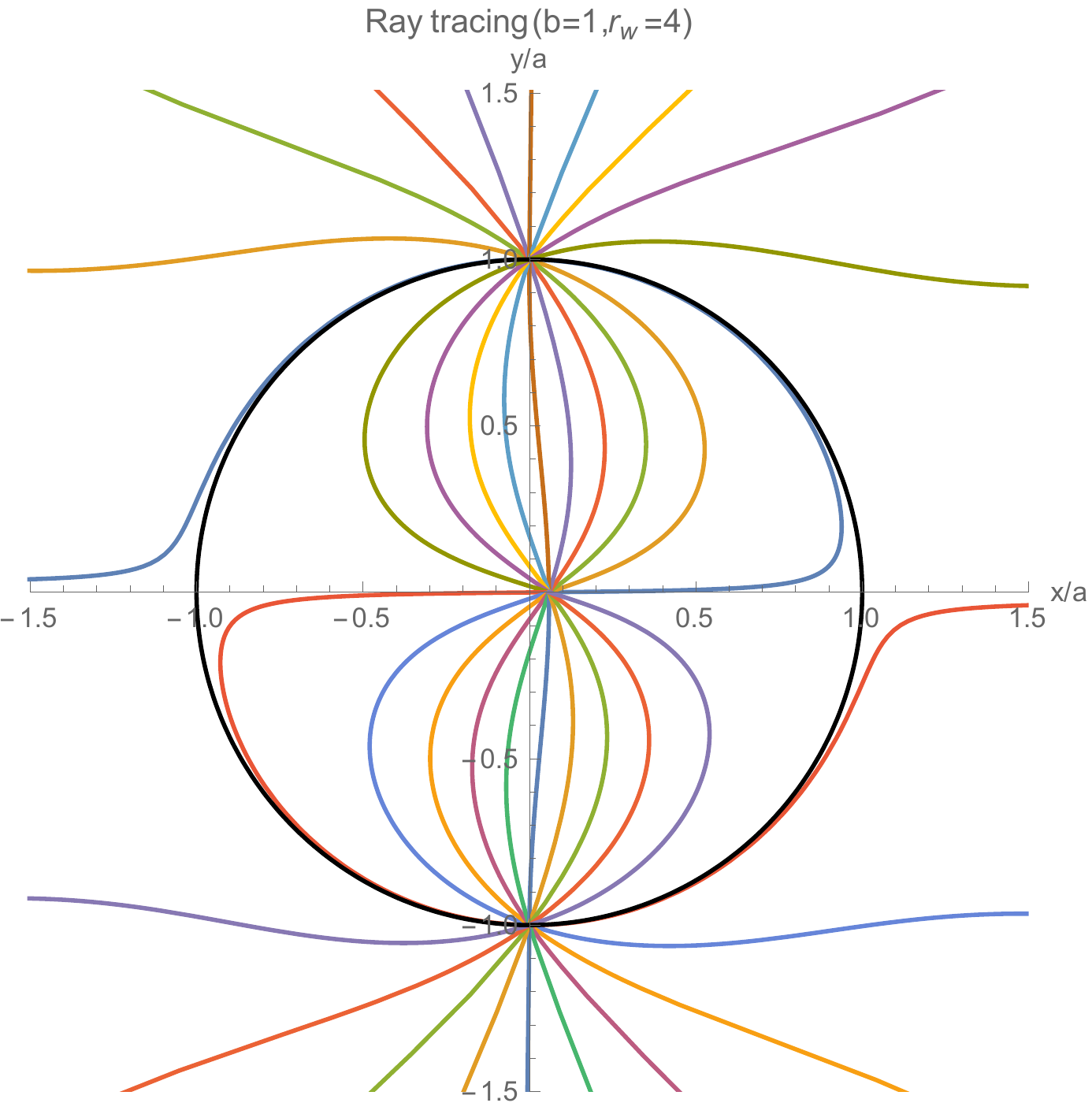}

\caption {\label{fig:ray} (Colour online) Ray trajectories computed according to Hamilton's equations. Unit circle is marked black to indicate the boundary between two Riemann sheets in virtual space. Rays starting with different directions are coloured differently for illustration. Parameters: $b=1,r_w=4$. }  
\end{figure}

Based on the refractive index distribution~\eref{nxy}, we use Hamilton's equations $\partial_t{\bold r}=\partial \omega/\partial \bold k$ and $\partial_t{\bold k}=-\partial \omega/\partial \bold r$ to compute the ray trajectories~\cite{Leonhardt2010book,Liu2013d} by \texttt{Mathematica}~\cite{Mathematica}, in which $\bold k$ indicates the momentum, $\bold r$ the position, $\omega$ the Hamiltonian. An instance is plotted in figure~\ref{fig:ray} for the same parameter as figure~\ref{fig:Ez} to illustrate the light flow of imaging: light rays originate from source point $z=0.0590$, diverge at two image points $z=\pm i$ and then propagate outward. This ray-tracing diagram aligns remarkably with the wave solutions we obtained in \fref{fig:Ez}.

\section{Summary}
In this manuscript, we demonstrate a new scheme to design heterogeneous optical medium(isotropic) from perspectives of wave simulation and analytic approximation, by making use of full Maxwell fisheye as well as conformal-mapping method~\cite{Leonhardt2006,Xu2014b}. The merit of this idea lies on removal of truncation of the profile of Maxwell fisheye, and that it maintains the same imaging resolution as a truncated Maxwell fisheye(with mirror). A possible alternative to compressing infinity into finiteness as conformal map does, could be the inverse tangent transform in Penrose diagrams(cf. pp201-3~\cite{Leonhardt2010book}). There is more versatile functionality to investigate on, if one puts on the lower sheet more than one light source, for instance the mutual interference between sources, whether constructively or destructively~\cite{Qiannan2013}.

\ack
The \texttt{COMSOL} Radio Frequency module simulation was performed mainly during L Y's 2013-2014 visit in Soochow University. L Y acknowledges the Singapore Ministry of Education Tier2 grant(MOE2008-T2-01-033) to sponsor his first trainings to learn \texttt{COMSOL} software in High Performance Computing Centre of Nanyang Technolgical University. L Y is sponsored by QUEST project grant EP/I034548/1. H C is supported by the National Science Foundation of China for Excellent Young Scientists (grant no. 61322504) and the National Excellent Doctoral Dissertation of China (grant no. 201217). L Y expresses gratitude to Prof. Yang Hao for his helpful suggestions and Zhao Junming for the assistance to plot \fref{fig:Fig1}(b).

\section*{Author contributions}
H C proposed this starting idea and co-designed it with L Y. L Y performed \texttt{COMSOL} simulation, completed all the computation, and prepared the manuscript draft. H C revised the manuscript.


\section*{References}

\bibliographystyle{unsrt}
	
\bibliography{bib5}









\end{document}